\documentclass[aps,prb,twocolumn,superscriptaddress]{revtex4-2}

\usepackage{amsmath, amssymb,  graphicx}

\usepackage[colorlinks=true ,urlcolor=blue,urlbordercolor={0 1 1}]{hyperref}

\usepackage{color}
\usepackage{xcolor}
\usepackage{braket}

\usepackage[utf8]{inputenc}
\usepackage{bm}
\usepackage{verbatim}
\usepackage{graphicx}
\usepackage{amsmath}
\usepackage{color}
\usepackage{float}
\usepackage{bbm}
\usepackage{amssymb}
\usepackage{slashed}
\usepackage{wasysym,bm,bbm,dsfont}

\begin{document}

\title{Symmetric topological Mott insulator and Mott semimetal}
\author{Boran Zhou}
\author{Ya-Hui Zhang}
\affiliation{Department of Physics and Astronomy, Johns Hopkins University, Baltimore, Maryland 21218, USA}

\date{\today}

\begin{abstract}
Correlated physics in nearly flat topological bands is a central theme in the study of moiré materials. While ground states at integer fillings are typically identified as quantum Hall ferromagnets within a Hartree-Fock framework, we propose the existence of symmetric topological Mott insulators (STMIs) that transcend this Slater determinant picture. Focusing on half-filling of each flavor per unit cell, we demonstrate the existence of STMIs which exhibit a quantized charge or spin Hall response. We first establish this phase in a bilayer Haldane-Hubbard model with localized orbitals on the $A$ sublattice and dispersive band on the $B$ sublattice. Starting from a trivial Mott insulator on the $A$ sublattice,  tuning the  sublattice potential drives a Bose–Einstein-condensation (BEC) to Bardeen–Cooper–Schrieffer (BCS) transition of the associated $p-\mathrm{i}p$ exciton pairing, realizing a topological Mott insulator with $C=1$ per flavor. We further generalize this construction to a single-layer spinful model, where the resulting STMI hosts charge edge modes coexisting with bulk local moments. A Mott semimetal is identified at the quantum critical point between the STMI and the trivial Mott insulator. Finally, we discuss applications to AA-stacked MoTe$_2$/WSe$_2$, proposing a ferromagnetic Chern insulator phase as a low-temperature descendant of the symmetric Mott semimetal.\end{abstract}

\maketitle
\textbf{Introduction}  The relation between strong correlations and band topology has long been a central theme in condensed matter physics~\cite{hohenadler2013correlation,rachel2018interacting}. In recent years, moir\'e materials have emerged as a powerful platform to explore this intersection, particularly regarding the nature of correlated states at integer fillings~\cite{nuckolls2024microscopic,andrei2021marvels,andrei2020graphene,mak2022semiconductor}. Unlike trivial bands which can be Wannierized and described by a standard Hubbard model, topological bands, such as those in twisted bilayer graphene (TBG)~\cite{cao2018magic,cao2018correlated,yankowitz2019tuning,lu2019superconductors,stepanov2020untying,cao2021nematicity,liu2021tuning,arora2020superconductivity} and  transition metal dichalcogenide (TMD) bilayer~\cite{zhao2024realization,li2021quantum,tao2024valley,wu2019topological,yu2020giant,devakul2021magic,pan2020band,zhang2021spin,cai2023signatures,zeng2023thermodynamic,park2023observation,xu2023observation}, typically resist localized Wannier orbitals. At integer fillings, the ground state is conventionally understood as a quantum Hall ferromagnet \cite{zhang2019nearly,repellin2020ferromagnetism} within the Hartree Fock framework. Extensive theoretical works have been developed under this paradigm~\cite{zhang2019twisted,PhysRevX.10.031034,PhysRevLett.122.246401,PhysRevX.11.041063,mai20231,zhao2023failure}, and supported by experiments~\cite{sharpe2019emergent,serlin2020intrinsic,bultinck2020mechanism,xie2021fractional}. However, Mott physics may be relevant even in these topological settings, suggesting physics beyond the mean field framework. This scenario is supported by experiments in TBG revealing signatures of local moments~\cite{rozen2021entropic,saito2021isospin} and ``Mottness'' alongside topological features. These observations have motivated proposals for nontrivial Mott phases compatible with the fragile topology of TBG~\cite{PhysRevLett.129.047601,PhysRevX.15.021087,zhao2025ancilla,zhao2025mixed,ledwith2025exotic,hu2025projected}.  In TBG, the Mott states are usually topologically trivial  due to the $C_2T$ symmetry. Then a natural question is: can we realize a symmetric topological Mott state with  quantum anomalous Hall (QAH) or quantum spin hall (QSH) effect in other systems?  


 In this paper, we study a model with four flavors (combining spin and layer/valley) or two spins at half filling of the band for each flavor. In the non-interacting case, the system must be in a Fermi liquid phase. We ask whether a Mott-like gap can be opened by strong interactions, akin to the Mott insulator with separated upper and lower Hubbard bands in Hubbard model. If the resulting state is a symmetric Chern insulator or quantum spin Hall (QSH) insulator, we call it symmetric topological Mott insulator (STMI). STMI clearly violates the perturbative Luttinger theorem which guarantees a  Fermi surface area of half  the Brillouin zone (BZ) per flavor. STMI is possible in the following two cases: (I) In bilayer model with inter-layer spin-spin coupling, but no inter-layer hopping, STMI is consistent with a non-perturbative Luttinger theorem~\cite{PhysRevLett.84.3370} and can be viewed as a topological version of the symmetric mass generation (SMG) insulator~\cite{wang2022symmetric} ; (II) In single layer model, STMI is possible at high temperature where local moments are fluctuating or at zero temperature when the local moments form a spin liquid with fractionalization.  We note that in the literature the notion ``topological Mott insulator" (TMI) has been used for different meanings, which we clarify in the following. Ref.~\cite{raghu2008topological,herbut2014topological,mai2023topological,chen2021realization} discuss TMI for symmetry breaking phases, while our focus here is on symmetric state beyond Slater determinant. The topological aspect of our theory is for the physical charged excitation, and is thus different from TMI in Ref.~\cite{wagner2024edge,pesin2010mott} focusing on the band topology of the neutral spinons and in Ref.~\cite{PhysRevResearch.6.033235,PhysRevB.108.125115} on zeros of the Green's function.  The TMIs in Ref.~\cite{TopCOMPRB,pasqua2025quasiparticle}  are closer to ours in spirit, but these studies focus on integer filling of the band for each flavor and hence the states there are still smoothly connected to a band insulator.

We start from a bilayer spinful Haldane-Hubbard model, where one sublattice $A$ hosts localized orbital $f_A$ and the other sublattice $B$ hosts itinerant electrons $c_B$. We introduce a sublattice potential difference $\Delta$ between $B$ and $A$ sites. In the large $\Delta$ limit, all electrons are localized on the A sublattice and we simply have a  trivial triangular-lattice Mott insulator: a strong inter-layer spin-spin coupling $J_A$ forces the local moments from $f_A$ to form rung singlets at filling $\nu=\frac{1}{2}$ per spin per layer.  Upon decreasing $\Delta$, we have equal densities of electrons and holes doped into the B and A sublattices.  The inter-sublattice hopping induces $p-\mathrm{i}p$ exciton pairing.  Increasing the exciton density tunes a Bose–Einstein-condensation (BEC) to Bardeen–Cooper–Schrieffer (BCS) transition, through which the trivial Mott insulator evolves into a topological Mott insulator with $C=1$ per flavor.  At the transition, a Mott semimetal emerges with a gap closing at $\Gamma$ point, while most momentum space hosts upper and lower Hubbard bands separated by a Mott gap at the scale of Hubbard $U$.

We  propose an explicit Gutzwiller-projected wavefunction for the STMI in the bilayer model, offering a concrete example of a Chern insulator or QSH insulator beyond Slater determinant. We also provide an equivalent description based on the ancilla theory~\cite{zhang2020pseudogap,zhou2025variational,zhao2025ancilla}. Then we generalize the description to the single layer  model at total filling $\nu_T=1$, where the local moments at the $A$ sublattice must be included in the effective theory. We assume there is a spin coupling $J_{\mathrm{eff}}$ between nearest neighbor AA sites. At finite temperatures $J_\mathrm{eff} \ll T \ll U$, the spin moments can be treated as thermally fluctuating, and we propose trivial MI to TMI transition tuned by sublattice potential $\Delta$ in the charge sector similar to the bilayer model. Especially there is a symmetric Mott semimetal in this intermediate temperature regime at the transition. As the temperature is lowered toward zero, we must decide the fate of the spin moments. When $J_{\mathrm{eff}}<0$, ferromagnetism emerges below a critical temperature $T_c$, which splits the single topological transition to two distinct topological transitions for the two spins.  In the intermediate $\Delta$, we have a ferromagnetic Chern insulator, which should be viewed as a descendant of the Mott semimetal above $T_c$. This picture may be relevant to the experiment in  MoTe$_2$/WSe$_2$ bilayer~\cite{li2021quantum}.

\begin{figure}[t]
    \centering
    \includegraphics[width=0.95\linewidth]{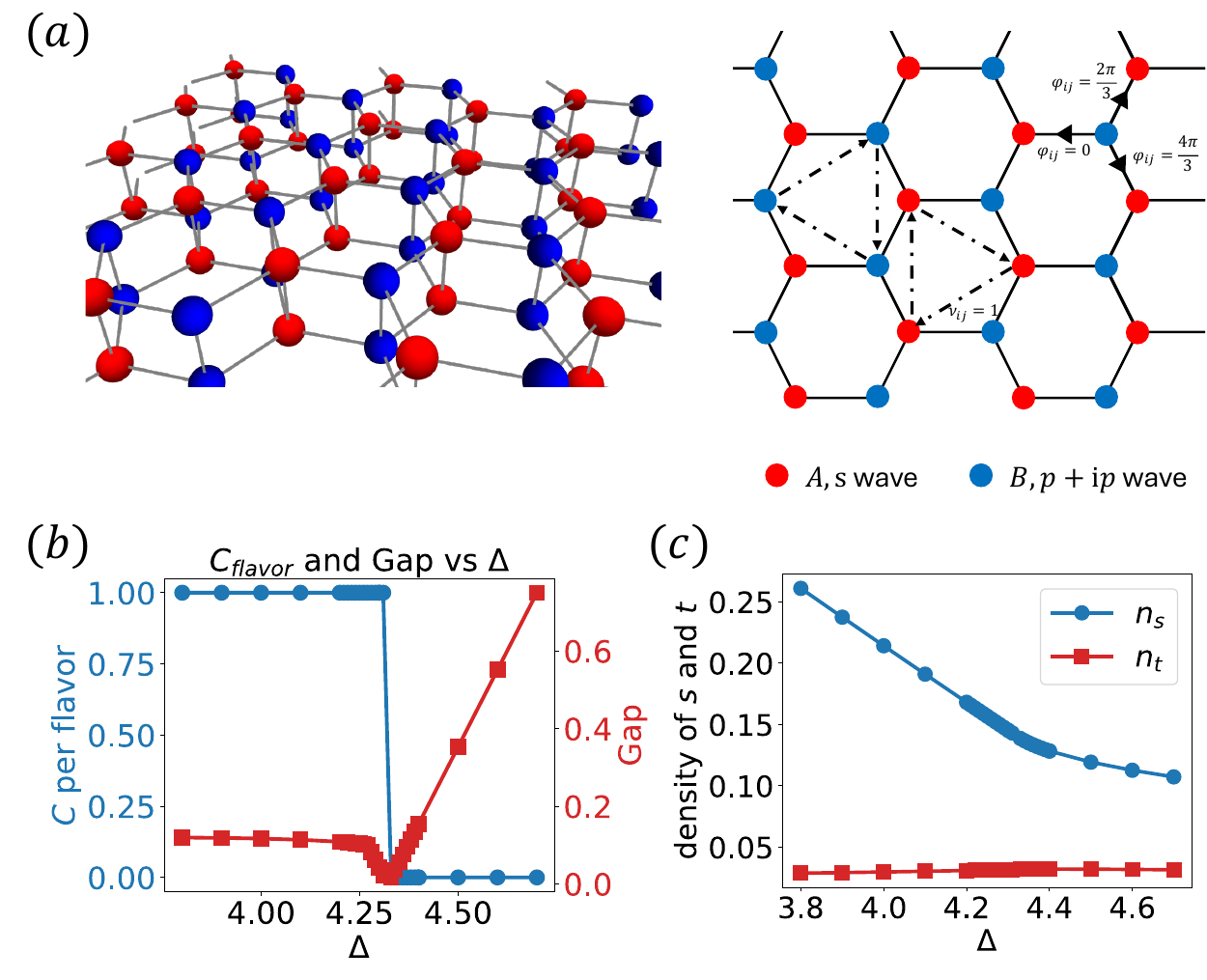}
    \caption{(a) Schematic illustration of the bilayer lattice model. For parameters $\gamma=1,\phi=\frac{\pi}{5},t_A=0.1,t_B=-2,U=2.5$: (b) Dependence of the Chern number (per flavor) and the many-body gap distribution on the sublattice potential difference $\Delta$. A topological transition occurs as $\Delta$ is increased. (c) Dependence of $n_s$ and $n_t$ on $\Delta$. In this regime, one finds $n_s\gg n_t\approx0$, indicating that the triplon excitations play a less important role.}
    \label{fig:model}
\end{figure}

\textbf{Model} We start from a bilayer spinful Haldane-Hubbard model:
\begin{equation}\label{eq:original_H}
\begin{split}
    H=&H_A+H_B+H_{AB}+\Delta(N_{B}-N_A)-\mu N\\
    &+\sum_{i_A}\left(J_AS_{A;t}\cdot S_{A;b}+\frac{U}{2}(n_{A;i}-2)^2\right),\\
    H_A=&\sum_{\braket{\braket{ij}}_A;\alpha;l}(t_Ae^{\mathrm{i}\nu_{ij}\phi}f^\dagger_{A;i;\alpha;l}f_{A;j;\alpha;l}+\mathrm{H.c.}),\\
    H_B=&\sum_{\braket{\braket{ij}}_B;\alpha;l}(t_Be^{\mathrm{i}\nu_{ij}\phi}c^\dagger_{B;i;\alpha;l}c_{B;j;\alpha;l}+\mathrm{H.c.}),\\
    H_{AB}=& \sum_{\braket{ij};\alpha;l}\gamma\left(e^{\mathrm{i}\varphi_{ij}}f^\dagger_{A;i;\alpha;l}c_{B;j;\alpha;l}+\mathrm{H.c.}\right),
\end{split}
\end{equation}
where $A$ is $s$ orbital and $B$ is $p+\mathrm{i}p$ orbital. $\alpha=\uparrow,\downarrow$ labels the spin and $l=t,b$ labels the layer. $\langle ij\rangle$ represents the nearest-neighbor bond. $\langle\langle ij\rangle\rangle_A$ and $\langle\langle ij\rangle\rangle_B$ represents the next-nearest-neighbor bond for sublattice $A$ and $B$. $\nu_{ij}=\pm1$, $\varphi_{ij}=0,\frac{2\pi}{3},\frac{4\pi}{3}$ depend on the path connecting $i$ and $j$, as illustrated in Fig.~\ref{fig:model}(a). We have continuous symmetry $\left(U(1)_c\times U(1)_v\times SU(2)_s\right)/Z_2$: $f_{A;i;\alpha;l},c_{B;i;\alpha;l}\rightarrow e^{\mathrm{i}(\theta_c+\theta_v\tau_z+\mathbf{\omega}\cdot\mathbf{\sigma})}f_{A;i;\alpha;l},e^{\mathrm{i}(\theta_c+\theta_v\tau_z+\mathbf{\omega}\cdot\mathbf{\sigma})}c_{B;i;\alpha;l}$. Here $\theta_c$ corresponds to the charge $U(1)_c$ rotation, $\theta_v$ corresponds to the layer $U(1)_v$ rotation generated by $\tau_z$ and $\mathbf{\omega}\cdot\mathbf{\sigma}$ represents the spin $SU(2)_s$ rotation. We also have a $C_3$ symmetry: $f_{A;i;\alpha;l},c_{B;i;\alpha;l}\rightarrow f_{A;C_{3}i;\alpha;l},e^{\mathrm{i}\frac{2\pi}{3}}c_{B;C_{3}i;\alpha;l}$.  In the following we assume a large $t_B$ and $\gamma$, but an almost vanishing $t_A$. The model thus hosts flat f orbitals on A which strongly hybridizes with  itinerant electrons on B, which is similar to the topological heavy fermion model (THFM) in twisted bilayer graphene\cite{PhysRevLett.129.047601}. But as we will see, the physics in this work has nothing to do with heavy fermion systems.

\textbf{Effective model in restricted Hilbert space}
We are interested in the regime of $\nu_T=2$, or half filling per flavor per unit cell. We start from the large $\Delta$ limit where all electrons are localized on the A sublattice. The most relevant local Hilbert space of $f_A$ is restricted to valence $f^{1+},f^{2+}$ and $f^{3+}$, which we refer to as the singlon, doublon and triplon states respectively. We construct an effective model within this restricted Hilbert space. For the doublon states, we only keep the spin singlet state due to the large anti-Hund coupling $J_A$. This state is written as $\ket{d_i}=\sum_\alpha \frac{\alpha}{\sqrt{2}}f^\dagger_{A;i;\alpha;t}f^\dagger_{A;i;\bar{\alpha};b}\ket{0}$. In this formulation, the $f_A$ orbital forms as a product state $\prod_i \ket{d_i}$ in the decoupling limit. Viewing this product state as the vacuum, the hybridization $c^\dagger_B f_A$ excites a subspace spanned by the $f^{1+}$ and $f^{3+}$ states. We define these states as $\ket{s_{i;\alpha;l}}=\sqrt{2}f_{A;i;\bar{\alpha};\bar{l}}\ket{d_i}$ and $\ket{t_{i;\alpha;l}}=\sqrt{2}f^\dagger_{A;i;\alpha;l}\ket{d_i}$. We then introduce two fermion operators as $s^\dagger_{i;\alpha;l}=\ket{s_{i;\alpha;l}}\bra{d_i}, t^\dagger_{i;\alpha;l}=\ket{t_{i;\alpha;l}}\bra{d_i}$. The corresponding density operators are defined as $n_{i;s}=\sum_{\alpha;l}\ket{s_{i;\alpha;l}}\bra{s_{i;\alpha;l}}$ and $n_{i;t}=\sum_{\alpha;l}\ket{t_{i;\alpha;l}}\bra{t_{i;\alpha;l}}$. The effective model requires that $n_{i;s}+n_{i;t}+\ket{d_i}\bra{d_i}=1$. Similar to the case of $t-J$ model and the previous work of topological heavy fermion model~\cite{zhao2025mixed}, the $f$ operator projected to this subspace takes the Gutzwiller-projected form:
\begin{equation}\label{eq:f_operator}
    P_G f^\dagger_{A;i;\alpha;l}P_G=\frac{1}{\sqrt{2}}\left(s_{i;\bar{\alpha};\bar{l}}+t^\dagger_{i;\alpha;l}\right).
\end{equation}
By replacing the original operators in Eq.~\ref{eq:original_H} with their projected forms, we obtain an effective model within the restricted Hilbert space. Specifically, we substitute $f_{A;i;\alpha;l}\rightarrow P_Gf_{A;i;\alpha;l}P_G$, and reformulate the on site $f$ terms to be the local energies of the $f^{1+},f^{2+},f^{3+}$ valence states. Then we relax the Gutzwiller projection $P_G$ and get the renormalized mean field Hamiltonian:
\begin{equation}\label{eq:meanfieldH}
\begin{split}
    H_\mathrm{MF}=&g^2_\gamma H_A +H_B+g_\gamma H_{AB} + \sum_{i\in A}\left(E_s n_{i;s}+E_tn_{i;t}\right)\\
    &+(\Delta-\mu) N_B,
\end{split}
\end{equation}
where $g_\gamma=\sqrt{1-\braket{n_s}-\braket{n_t}}, E_s=\Delta+\frac{U}{2}+\frac{3J_A}{4}+\mu,E_t=-\Delta+\frac{U}{2}+\frac{3J_A}{4}-\mu$. In our calculation, we assume that $J_A\ll U$, therefore $E_s$ and $E_t$ can be simplified as $E_s=\Delta+\frac{U}{2}+\mu$, $E_t=-\Delta+\frac{U}{2}-\mu$. The ground state is $\ket{\Psi}=P_G \ket{\mathrm{Gauss}[s,t,c_B]}$, where $\ket{\mathrm{Gauss}[s,t,c_B]}$ is the ground state of the mean field Hamiltonian Eq.~\ref{eq:meanfieldH}. $P_G$ enforces that $n_{i;s}+n_{i;t}\le1$ for each sublattice.

\textbf{Topological Mott insulator} Based on the mean field Hamiltonian introduced above, we self-consistently obtain $g_\gamma$ and get the ground state wavefunction. The resulting phase evolution is shown in Fig.~\ref{fig:model}(b). In the calculation, we fix the model parameters to be
$\gamma = 1, \phi = \frac{\pi}{5}, t_A = 0.1, t_B = -2$ and $ U = 2.5$.
By gradually increasing  $\Delta$, we observe a topological phase transition characterized by a change in the Chern number per flavor from $C_{\mathrm{flavor}} = 1$ to $C_{\mathrm{flavor}} = 0$. The $C_{\mathrm{flavor}} = 1$ and $C_{\mathrm{flavor}}=0$ phase corresponds to the topological Mott insulator and trivial Mott insulator respectively.

We find that triplon excitations play only a minor role in this parameter regime. As shown in Fig.~\ref{fig:model}(c), the singlon density is much larger than the triplon density, $n_s\gg n_t$, indicating that the low energy fluctuations are dominated by the $f^{1+}$ sector. Furthermore, in the topological Mott insulator phase, the momentum distribution $n_s(\mathbf{k})$ mainly concentrates in a small region around $\mathbf{k}=0$ in the Brillouin zone. This behavior suggests that the nontrivial topology originates from the mixed-valence nature near the $\Gamma$ point.

\begin{figure}[t]
    \centering
    \includegraphics[width=0.95\linewidth]{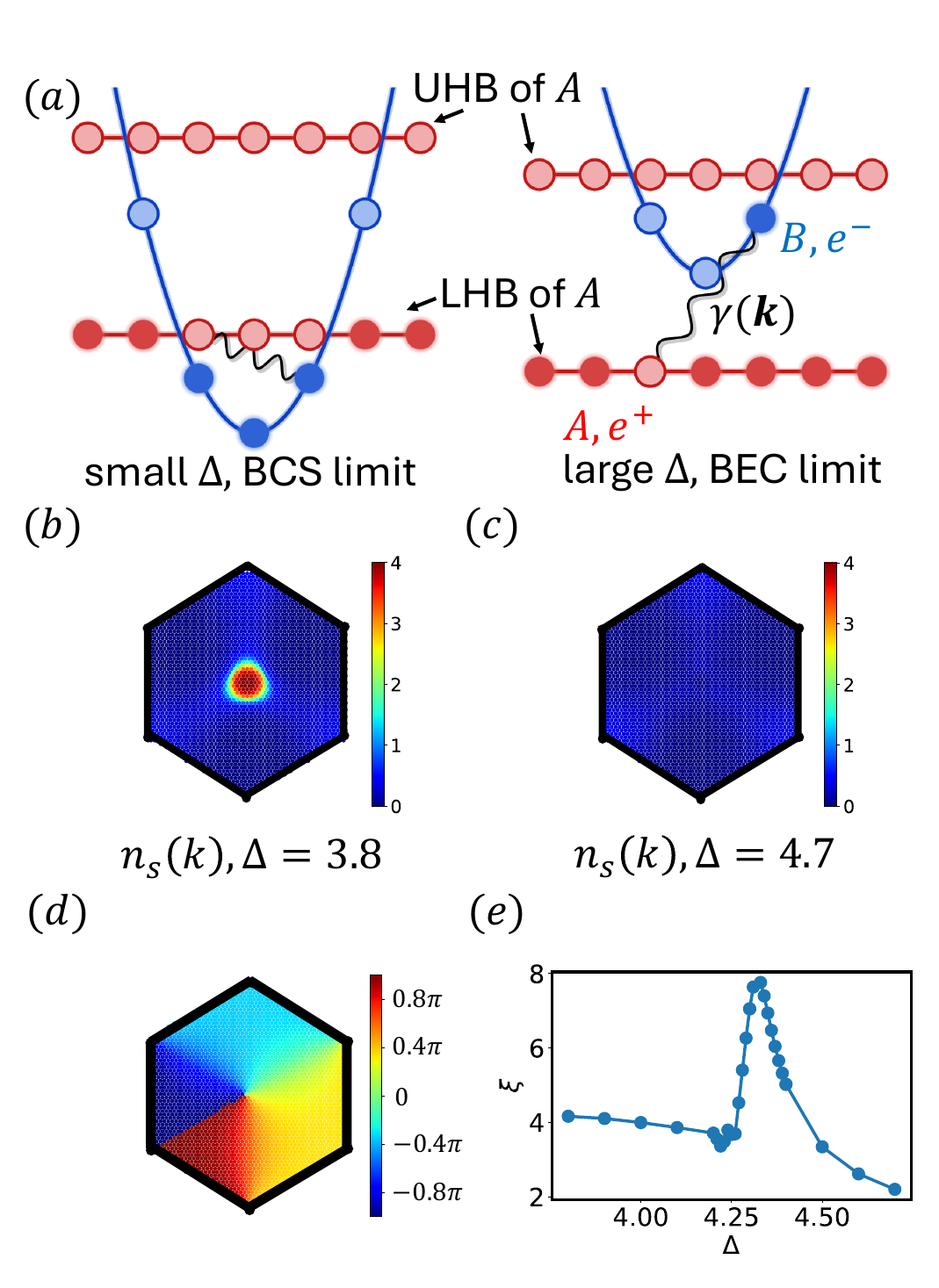}
    \caption{(a) Schematic illustration of the exciton pairing in the small $\Delta$ (BCS) and large $\Delta$ (BEC) regimes respectively. For small $\Delta$, the system is in the BCS regime.  For large $\Delta$, the energy cost of exciton is high, leading to a tightly bound BEC regime. For parameters $\gamma=1,\phi=\frac{\pi}{5},t_A=0.1,t_B=-2,U=2.5$: (b) and (c) show the momentum density distribution $n_s(\mathbf{k})$ for $\Delta=3.8$ and $4.7$ respectively. At $\Delta=3.8$, the system lies in the topological Mott insulator phase and $n_s(\mathbf{k})$ is sharply peaked around the $\Gamma$ point of the Brillouin zone. (d) Phase of the exciton pairing $\langle c^\dagger_{B;\alpha;l}(\mathbf{k})s_{\bar{\alpha};\bar{l}}^\dagger(-\mathbf{k})\rangle$ per flavor, which maintains a $p-\mathrm{i}p$ structure in the whole range of $\Delta$.  (e) Dependence of the exciton correlation length on $\Delta$. The correlation length diverges at transition point, indicating a continuous phase transition between the BEC and BCS regimes. }

    \label{fig:BEC_to_BCS}
\end{figure}

\textbf{BEC-to-BCS transition of exciton} We have concluded that the self doped singlon state on $A$ and electron on $B$ is the dominant reason to produce the nontrivial  topology. Therefore we can further simplify our model to exclude $t_{i;\alpha;l}$ operators. After this simplification, the hybridization term $c_{B;\alpha;l}^\dagger f_{A;\alpha;l}$ takes the form of a pairing operator $c^\dagger_{B;\alpha;l} s^\dagger_{\bar{\alpha};\bar{l}}$. The physical meaning of $\langle c^\dagger_{B;\alpha;l} s^\dagger_{\bar{\alpha};\bar{l}}\rangle$ can be interpreted as the exciton pairing between a hole on the $A$ sublattice and an electron on the $B$ sublattice. We plot $\braket{c^\dagger_{B;\alpha;l}(\mathbf{k})s^\dagger_{\bar{\alpha};\bar{l}}(-\mathbf{k})}$ in Fig.~\ref{fig:BEC_to_BCS}(d), which is  in a $p-\mathrm{i}p$ pairing around $\Gamma$ point. The origin of this pairing symmetry is that the $c_B$ and $f_A$ orbitals carry different angular momenta at $\mathbf{k}=0$, enforcing the $p$ wave nature of the exciton. 

The topological transition can be understood as strong to weak pairing of  the excitons.  In the trivial Mott insulator phase, the energy cost of creating an exciton pair is large due to the large sublattice potential $\Delta$, therefore the exciton density is small and we are on the BEC side.  Decreasing $\Delta$ drives the system into a BCS-like phase, where the excitons are more weakly bound and inherit nontrivial topology from the underlying band structure ~\cite{PhysRevB.61.10267}. The difference is shown in Fig.~\ref{fig:BEC_to_BCS}(a). Consistent with a topological phase transition, the correlation length of the exciton pairing diverges at the critical point, as shown in Fig.~\ref{fig:BEC_to_BCS}(e). On the BCS side, each flavor provides a Chern number $C=1$, resulting in a $C=4$ Chern insulator.


\begin{figure}[t]
    \centering
    \includegraphics[width=0.95\linewidth]{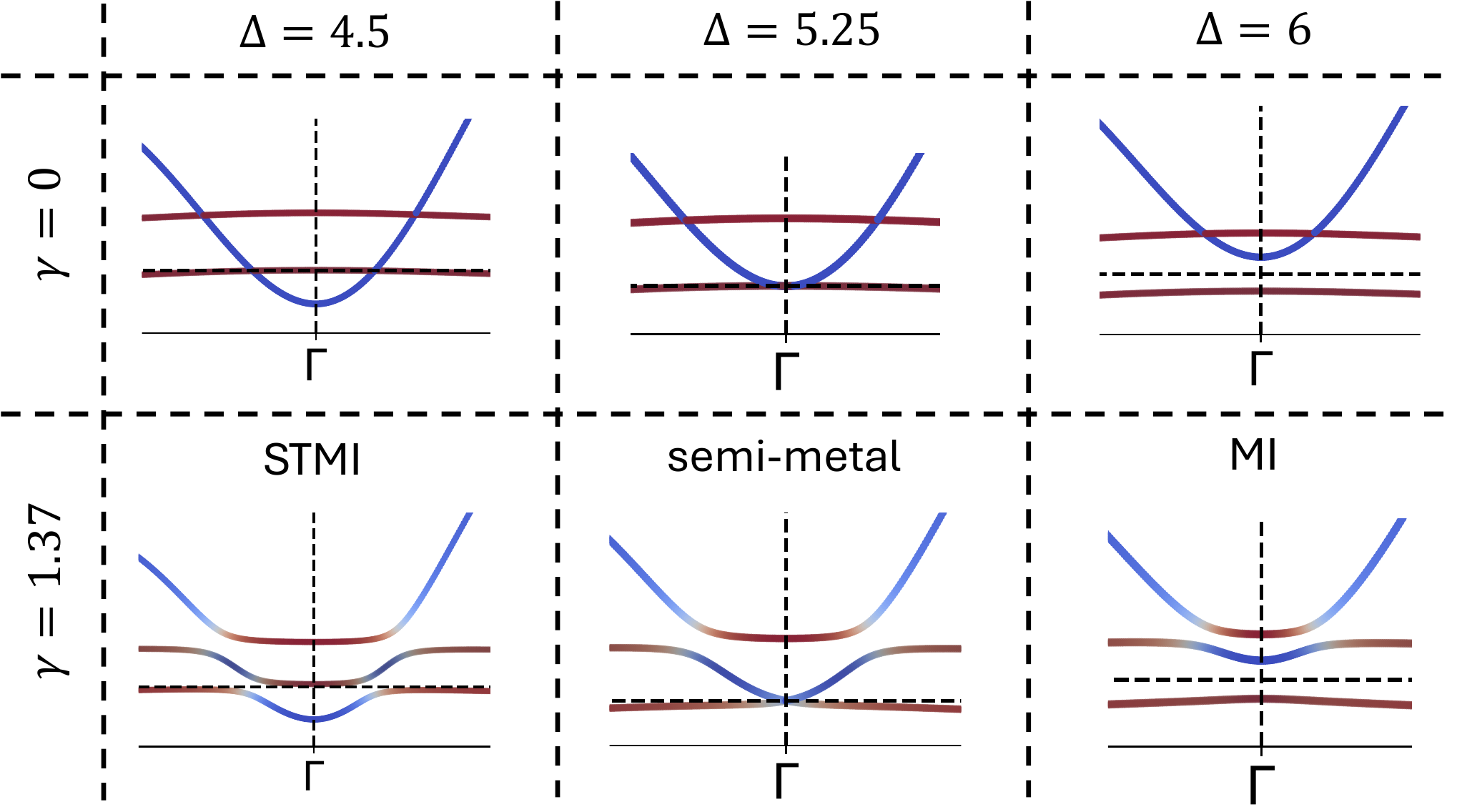}
    \caption{Parameters $\phi=0.78\pi,t_A=-0.1,t_B=2.5,U=2.5$. Band structures per flavor of the effective Hubbard model for various sublattice potential $\Delta$ and inter-sublattice hybridizations $\gamma$. Top row: $\gamma = 0$, bottom row: $\gamma = 1.37$. For large $\Delta$, the system is a trivial Mott insulator. For small $\Delta$ and finite $\gamma$, the system is a STMI, characterized by band inversion at $\Gamma$ point. The transition between STMI and MI is described by a Dirac cone (per flavor),  as shown in the middle panel.}
    \label{fig:table}
\end{figure}

\textbf{Topological Hubbard bands} Our mean field theory in Eq.~\ref{eq:meanfieldH} also provides a good description of the Hubbard bands. While Eq.~\ref{eq:f_operator} defines the physical $f_A$ operators as a linear combination of $s$ and $t^\dagger$, it is useful to introduce a complementary orthogonal operators:
\begin{equation}\label{eq:psi_oprator}
    \psi^\dagger_{i;\alpha;l}=\frac{1}{\sqrt{2}}\left(-s_{i;\bar{\alpha};\bar{l}}+t^\dagger_{i;\alpha;l}\right).
\end{equation}
The $\psi$ operators share the same symmetry transformation as the $f_A$ operators. By expressing Eq.~\ref{eq:meanfieldH} in terms of $f_A$ and $\psi$, we obtain an effective band theory for the Hubbard bands. In this picture, the interaction $U$ hybridizes the physical $f_A$ orbitals with the auxiliary $\psi$ fermions, splitting the upper and lower Hubbard bands for $A$ sublattice.  When $\Delta$ is reduced, the dispersive $c_B$ band touches and hybridizes with the lower Hubbard band (LHB) and a band inversion happens at the $\Gamma$ point when $\Delta<\Delta_c$.The inter-sublattice hybridization $\gamma$ opens an energy gap at this crossing, giving rise to a non-zero Chern number per flavor in the system, as shown in Fig.~\ref{fig:table}.



\textbf{Equivalence to the ancilla theory} We now show that the renormalized mean field theory is equivalent to the ancilla formulation. In the ancilla approach\cite{zhang2020pseudogap,zhao2025ancilla}, two auxiliary fermionic layers $\psi$ and $\psi^\prime$ are introduced on the $A$ sublattice. In this formalism, the physical fermions $c_B,f_A$ and the $\psi$ fermion together form the charge sector,  while the $\psi^\prime$ form as the charge neutral spin sector. The final ancilla wavefunction can be written as:
\begin{equation}
    \ket{\Psi_\mathrm{ancilla}}=P_S \ket{\mathrm{Slater}[c_B,f_A,\psi]}\otimes\ket{\Psi_{\psi^\prime}}.
\end{equation}
where $P_S$ enforces that the
two ancilla fermions form an $SU(N_\mathrm{flavor})$ spin singlet at each site
$i$, where $N_\mathrm{
flavor
}=4$ in the bilayer example. The state $\ket{\psi^\prime}$ is the spin wavefunction when $U\rightarrow\infty$, while $\ket{\mathrm{Slater}[c_B,f_A,\psi]}$ is determined by the mean field Hamiltonian for the charge sector:
\begin{equation}\label{eqn:Hancilla}
\begin{split}
    H_\mathrm{ancilla}=&H_A+H_B+H_{AB}+\Delta\left(N_B-N_A\right)-\mu N\\
    &+\Phi\sum_{\mathbf{k};\alpha;l}\left( f^\dagger_{A;\mathbf{k};\alpha;l}\psi_{\mathbf{k};\alpha;l}+\mathrm{H.c.}\right)-\mu_\psi N_\psi,
\end{split}
\end{equation}
where the chemical potential $\mu_\psi$ is tuned to fix the constraint $\braket{n_{i;\psi}}=N_\mathrm{flavor}-\nu_T$ for each site. Note here $\nu_T=2$. After expressing $f_A$ and $\psi$ in terms of the singlon and triplon operators $s,t$ as $f^\dagger_{A;\alpha;l}=\frac{1}{\sqrt{2}}(s_{\bar{\alpha};\bar{l}}+t^\dagger_{\alpha;l}),\psi^\dagger_{\alpha;l}=\frac{1}{\sqrt{2}}(-s_{\bar{\alpha};\bar{l}}+t^\dagger_{\alpha;l})$, which is exactly same as Eq.~\ref{eq:f_operator},\ref{eq:psi_oprator}, Eq.~\ref{eqn:Hancilla} maps directly onto the renormalized mean field Hamiltonian in Eq.~\ref{eq:meanfieldH}.  In this formulation, the spin sector described by $\psi'$ is detached from the charge sector. For the bilayer model, $\psi'$ is simply in a rung singlet phase, and thus can be ignored. In the previous treatment, a large $J_A$ is required. In the ancilla framework, we can see that the charge sector remains the same at smaller $J_A$, as long as the localized moments $\psi'$ are gapped.

\textbf{Mott semimetal at critical point} For the mean field Hamiltonian in Eq.~\ref{eqn:Hancilla}, as the parameter $\Delta$ is tuned to the critical value $\Delta_c$, the system enters a Mott semimetal phase described by the following effective Hamiltonian for each spin:
\begin{equation}\label{eq:effectiveH}
    H_\mathrm{eff}=v_\mathrm{eff}\left(-k_x\rho_y+k_y\rho_x\right)+\left(\Delta-\Delta_c \right)\rho_z,
\end{equation}
where $\rho$ is the Pauli matrix in the space of $(c_{B;\alpha;l},\frac{f_{A;\alpha;l}-\psi_{\alpha;l}}{\sqrt{2}})$, $v_\mathrm{eff}$ is the effective velocity.  Basically the dispersive $c_{B}$ band and the lower Hubbard band form a Dirac cone together, as shown in the middle plot of  Fig.~\ref{fig:table}.  

\textbf{Single layer Model}  We now attempt to generalize the STMI to a  single layer model: we simply remove the layer index $l$ and the $J_A$ term in Eq.~\ref{eq:original_H}.
Now the spin fluctuation becomes relevant at low energy and the renormalized mean field approach is no longer adequate. In contrast, the ancilla approach  can be easily generalized to the single layer model.

\begin{figure}[t]
    \centering
    \includegraphics[width=0.95\linewidth]{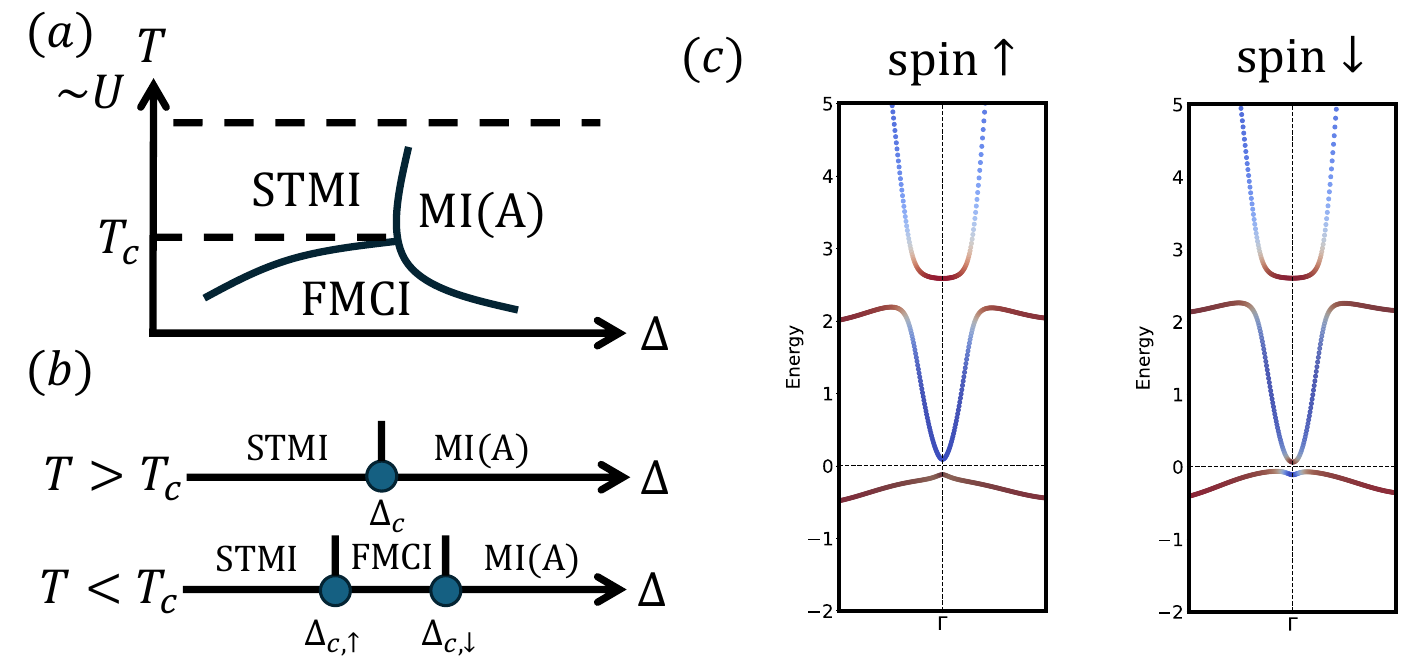}
    \caption{(a) Schematic phase diagram of the single layer model. We target the regime that $T\ll U$. At high temperature and large $\Delta$, the system is a triangular-lattice Mott insulator on $A$ sublattice, labeled as MI(A).It transits to the symmetric topological Mott insulator (STMI) through a symmetric Mott semimetal at the critical point. At lower $T$, the spin moments on A sublattice develop a ferromagnetic order, which splits the topological transition for the two spins, leading to an intermediate  ferromagnetic Chern insulator (FMCI) phase. (b) Linecut showing the phase boundary for $T>T_c$ and $T<T_c$ respectively. For $T<T_c$,  Spontaneous symmetry breaking generates an effective Zeeman field that lifts spin degeneracy, resulting in distinct critical values $\Delta_c$ for spin $\uparrow$ and spin $\downarrow$ sectors. (c) At critical point $\Delta=\Delta_c$, the ferromagnetic ordering drives the system into a  Chern insulator phase. Here we show the Hubbard band plot for spin $\uparrow$ and $\downarrow$ respectively in the ancilla framework. The parameters for the plot are: $\gamma=1,\phi=0.78\pi,t_A=-0.1,t_B=2.5,\Phi=1.25,\Delta=5.18,B=0.1,B^\prime=0.2$. Red (blue) color represents $A(B)$ sublattice.}
    \label{fig:1layer}
\end{figure}

Based on this ancilla framework, we conjecture a schematic phase diagram shown in Fig.~\ref{fig:1layer}(a) at total filling $\nu_T=1$ per unit cell, or half filling per spin. We focus on the regime $T\ll U$, where  the charge sector ($c_B,f_A,\psi$) and the spin sector ($\psi^\prime$) are separated.  $\psi'$ here represents the local moments on A sublattice and there is an inter-site effective spin coupling $J_\mathrm{eff}$ which governs its behavior at low temperature.  However, in the regime $|J_{\mathrm{eff}}|\ll T\ll U$, we expect that the local moments are thermally fluctuating and detached from the charge sector. In this regime, the charge sector is described by the same mean field Hamiltonian as in the bilayer model and we again expect a trivial Mott insulator to STMI transition by reducing the potential $\Delta$.  In the STMI phase,  there is a $C=2$ QAH effect coexisting with the bulk spin moments.  Our construction can also be generalized to the Kane-Mele Hubbard model, where the two spins have opposite Chern number in the non-interacting band. In this case, the STMI phase hosts QSH effect together with fluctuating spin moments. The spin Chern number is 2, which is the same as a QSH band insulator at $\nu_T=2$. We note that Ref.~\cite{mai20231} also discussed a QSH state at the same filling in a similar model, but the spin Chern number is half of ours. Unlike the usual QAH or QSH insulator, the fluctuating bulk spin moments should contribute to a Curie-Weiss law for spin susceptibility, which may be experimentally confirmed. 

At lower temperature with $T\lesssim |J_\mathrm{eff}|$, the spin sector may become unstable towards ordering. Phenomenologically we expect a ferromagnetic $J_{\mathrm{eff}}$ when $\Delta$ is small. Let us now assume the two spins carry opposite Chern numbers as in moir\'e systems based on TMD.  To capture ferromagnetic (FM) ordering below a critical temperature $T_c \sim |J_{\mathrm{eff}}|$, we introduce an effective Zeeman field taking the form $B(N_{A;\uparrow}+N_{B;\uparrow}-N_{A;\downarrow}-N_{B;\downarrow})-B^\prime(N_{\psi\uparrow}-N_{\psi\downarrow})$ to the effective Hamiltonian in Eq.~\ref{eq:effectiveH}(see Supplemental Material). Substituting this term into the effective Hamiltonian, we find that the critical point for the topological transition splits for the two spin sectors: $\Delta_{c,\uparrow/\downarrow}=\Delta_{c,0}\mp\frac{B+B^\prime}{2}$, as illustrated in Fig.~\ref{fig:1layer}(b). In the intermediate $\Delta$, we have a FM Chern insulator with $C=1$. We note that spin is not necessarily fully polarized for the QAH effect in our description. Instead, the QAH insulator may be better viewed as a descendant of the symmetric Mott semimetal from adding a small FM order. Lastly, we note that a symmetric STMI phase even at $T=0$ is possible if the spin moments form a spin liquid. We provide an explicit wavefunction in the appendix to demonstrate the proof of existence in Appendix.~\ref{appendix:calculation_for_single_layer}.

  \textbf{Discussion} Ferromagnetic Chern insulator has been experimentally observed in MoTe$_2$/WSe$_2$ bilayer close to the charge transfer transition from a trivial Mott insulator~\cite{li2021quantum}. Our theory provides a natural description of the system by viewing the MoTe$_2$ and WSe$_2$ layers as the A and B sublattices in our spinful model.  It is interesting to search for possible trivial Mott insulator to STMI transition and symmetric Mott semimetal at higher temperature, as is illustrated in Fig.~\ref{fig:1layer}.  Note our theory applies to the AA stacking, and is thus different from theories developed for AB stacked MoTe$_2$/WSe$_2$~\cite{PhysRevX.12.021031,PhysRevB.107.L081101,PhysRevLett.129.056804,zhang2021spin}. 

\textbf{Conclusion} In summary, we have demonstrated the existence of a symmetric topological Mott insulator (STMI) at the half-filling of a topological band. Unlike standard topological insulators, these states—which host QAH or QSH effects—cannot be captured within a Slater determinant framework. We constructed an explicit STMI phase on a honeycomb lattice composed of a flat band on the A sublattice and a dispersive band on the B sublattice. The underlying mechanism is identified as inter-sublattice $p-\mathrm{i}p$ exciton pairing. The quantum phase transition between the trivial Mott insulator and the STMI is marked by the emergence of a Mott semimetal with a single Dirac cone per flavor, reminiscent of the physics discussed at charge neutrality of  twisted bilayer graphene~\cite{PhysRevX.15.021087,zhao2025ancilla}. These results may be relevant for moir\'e systems such as twisted bilayer graphene and twisted WSe$_2$~\cite{xia2025superconductivity,guo2025superconductivity}, where localized orbitals coexist with dispersive bands~\cite{PhysRevLett.129.047601,kim2025theory,xie2025superconductivity,crepel2024bridging}. Crucially, our analysis shows that at integer filling, the heavy fermion paradigm is inapplicable; instead, topological exciton pairing provides the correct physical picture for these mixed-valence Mott states.

\textbf{Acknowledgement} We thank Jing-Yu Zhao for previous collaborations. The work is supported by the Alfred P. Sloan Foundation through a Sloan Research
Fellowship (Y.-H.Z.).

\bibliographystyle{apsrev4-1}
\bibliography{main}

\appendix
\section{Ancilla theory of STMI}\label{appendix:ancilla_theory}
\begin{figure}[t]
    \centering
    \includegraphics[width=0.95\linewidth]{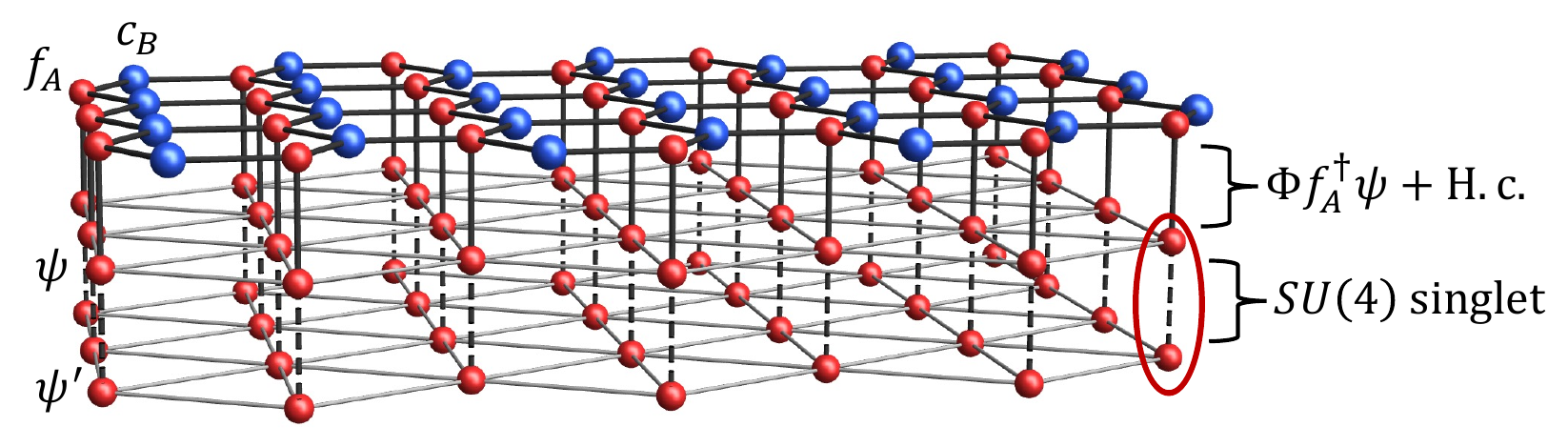}
    \caption{Illustration of the ancilla wave function. The blue circle represents the itinerant electron $c_B$, while the red circle contains the local fermion $f_A$ and two ancilla fermions, $\psi$ abnd $\psi^\prime$. The on-site hybridization $\Phi$ couples $f_A$ and $\psi$, leaving $\psi^\prime$ decoupled from the other system. Here, we treat the layer index as a flavor index rather than explicitly showing a physical bilayer structure.}
    \label{fig:ancilla}
\end{figure}

In this section, we briefly review the ancilla theory and its application to the bilayer spinful Haldane-Hubbard model introduced in the main text. In the ancilla theory, we  introduce two ancilla fermions $\psi$ and $\psi^\prime$, considering the following ansatz:
\begin{equation}\label{eq:ancilla_wavefunction}
    \ket{\Psi_\mathrm{ancilla}}=P_S\left(\ket{\mathrm{Slater}[c_B,f_A,\psi]}\otimes\ket{\Psi_{\psi^\prime}}\right),
\end{equation}
where $P_S$ is a projection operator enforcing: (I) $n_{i;\psi}=4-\nu_T$; (II) $n_{i;\psi^\prime}=\nu_T$; (III) the two local ancilla qubits $\psi_i$ and $\psi^\prime_i$ form as $SU(4)$ singlet at each site $i$. The construction of ancilla wavefunction is illustrated in Fig.~\ref{fig:ancilla}. Here we target the filling $\nu_T=2$ and choose the wavefunction of $\psi^\prime$ to be:
\begin{equation}
    \ket{\Psi_{\psi^\prime}}=\prod_{i=1}^{N_\mathrm{site}}\frac{1}{\sqrt{2}}\left(\psi^{\prime\dagger}_{i;\uparrow;t}\psi^{\prime\dagger}_{i;\downarrow;b}-\psi^{\prime\dagger}_{i;\downarrow;t}\psi^{\prime\dagger}_{i;\uparrow;b}\right)\ket{0}
\end{equation}
due to large anti-Hund's coupling $J_A$. The charge sector is determined by the physical fermions $c_B$, $f_A$ and the first ancilla fermion $\psi$. This is taken as a Slater determinant $\ket{\mathrm{Slater}[c_B,f_A,\psi]}$ and determined by the following mean field Hamiltonian:
\begin{equation}
\begin{split}
H_\mathrm{ancilla}=&H_A+H_B+H_{AB}+\Delta\left(N_B-N_A\right)-\mu N\\
    &+\Phi\sum_{\mathbf{k};\alpha;l}\left( f^\dagger_{A;\mathbf{k};\alpha;l}\psi_{\mathbf{k};\alpha;l}+\mathrm{H.c.}\right)-\mu_\psi N_\psi,
\end{split}
\end{equation}
where $\mu$ and $\mu_\psi$ are tuned to make $\braket{N_A}+\braket{N_B}=\nu_T N_{\mathrm{site}}$ and $\braket{N_\psi}=(4-\nu_T)N_{\mathrm{site}}$.

\begin{figure}[t]
    \centering
    \includegraphics[width=0.95\linewidth]{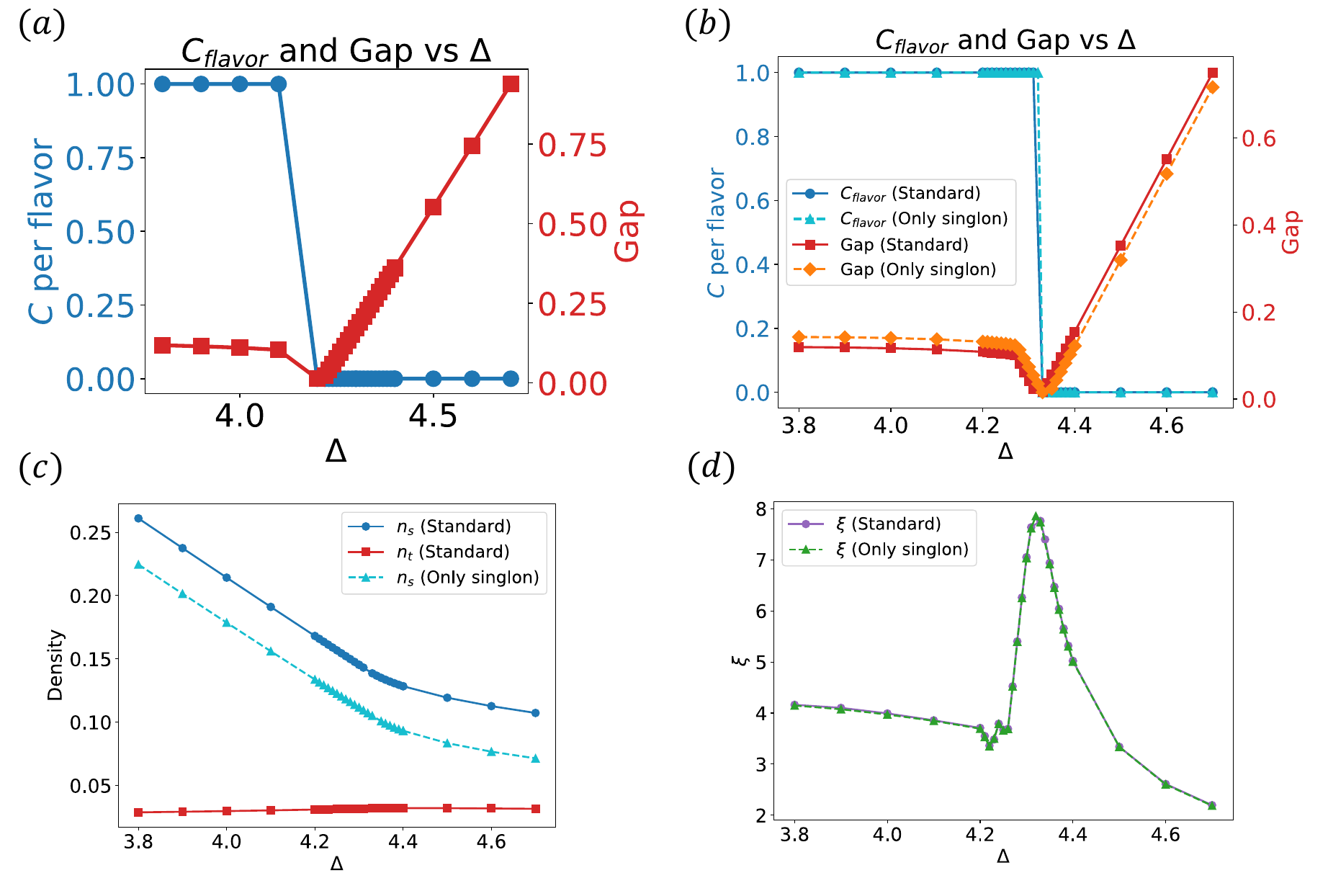}
    \caption{(a)(b) Dependence of the Chern number per flavor ($C_{\text{flavor}}$, left axis) and the gap (right axis) on sublattice potential $\Delta$. In (a), the result is obtained by implementing ancilla theory for parameters $\gamma=1,\phi=\frac{\pi}{5},t_A=0.1,t_B=-2,\Phi=1.25.$ In (b), the result is obtained by using renormalized mean field theory, comparing the full formalism (including $f^{3+}$ states) and the restricted description (excluding $f^{3+}$ states). The parameters are $\gamma=1,\phi=\frac{\pi}{5},t_A=0.1,t_B=-2,U=2.5.$
(c) Dependence of $n_s$ and $n_t$ on $\Delta$. (d) Dependence of the exciton correlation length on $\Delta$. Panels (c) and (d) follow the same parameter set and projection scheme as in (b). }
    \label{fig:appx1}
\end{figure}

We implement the ancilla theory and perform the calculation for the charge sector, obtaining the dependence of gap and Chern number per flavor as shown in Fig.~\ref{fig:appx1}(a). It has a good agreement with the result in Fig.~1(b) in the main text.

\section{Calculation excluding triplons}
In the main text, we perform our calculations using the renormalized mean field theory including $f^{1+},f^{2+},f^{3+}$ states. In this section, we perform the same calculation excluding $f^{3+}$ states to test the importance of the triplon excitations. Now the mean field Hamiltonian is written as:
\begin{equation}
\begin{split}
    H^\prime_\mathrm{MF}=&g^{\prime2}_\gamma H^\prime_A +H_B+g^\prime_\gamma H^\prime_{AB} + \sum_{i\in A}E_s n_{i;s}\\
    &+(\Delta-\mu)N_B, \\
    H^\prime_A=&\sum_{\braket{\braket{ij}}_A;\alpha;l}(\frac{t_A}{2}e^{\mathrm{i}\nu_{ij}\phi}s_{A;i;\bar{\alpha};\bar{l}}s^\dagger_{A;j;\bar{\alpha};\bar{l}}+\mathrm{H.c.}),\\
    H_B=&\sum_{\braket{\braket{ij}}_B;\alpha;l}(t_Be^{\mathrm{i}\nu_{ij}\phi}c^\dagger_{B;i;\alpha;l}c_{B;j;\alpha;l}+\mathrm{H.c.}),\\
    H^\prime_{AB}=& \sum_{\braket{ij};\alpha;l}\frac{\gamma}{\sqrt{2}}\left(e^{\mathrm{i}\varphi_{ij}}s_{A;i;\bar{\alpha};\bar{l}}c_{B;j;\alpha;l}+\mathrm{H.c.}\right),
\end{split}
\end{equation}
where $g^\prime_\gamma=\sqrt{1-\braket{n_s}}$ and $E_s=\Delta+\frac{U}{2}$, $\nu_{ij}$ and $\varphi_{ij}$ are defined same as Eq.~1 in the main text. The self consistent calculation result is shown in Fig.~\ref{fig:appx1}(b)-(d), which is in a good agreement with the results including $f^{3+}$ states.
\section{Single layer Haldane-Hubbard model calculation}\label{appendix:calculation_for_single_layer}
In this section, we provide the details about the calculations for the single layer Haldane-Hubbard model calculation. For single layer, spin fluctuations become significant at low energies, necessitating an ancilla approach that explicitly manifests spin-charge separation. The ancilla mean field Hamiltonian takes the form:
\begin{equation}
\begin{split}
H_\mathrm{ancilla}=&H_A+H_B+H_{AB}+\Delta\left(N_B-N_A\right)-\mu N\\
    &+\Phi\sum_{\mathbf{k};\alpha}\left( f^\dagger_{A;\mathbf{k};\alpha}\psi_{\mathbf{k};\alpha}+\mathrm{H.c.}\right)-\mu_\psi N_\psi\\
    &+J_K\sum_{i\in A}\left(\braket{\mathbf{S}_{i;A}}\cdot \mathbf{S}_{i;\psi}+\braket{\mathbf{S}_{i;\psi}}\cdot\mathbf{S}_{i;A}\right)\\
&+J_\perp\sum_{i\in A}\braket{\mathbf{S}_{\psi^\prime}}\cdot \mathbf{S}_\psi,
\end{split}
\end{equation}
where $\mu$ and $\mu_\psi$ are tuned to make $\braket{N_A}+\braket{N_B}=\nu_T N_{\mathrm{site}}$ and $\braket{N_\psi}=(2-\nu_T)N_{\mathrm{site}}$. The terms $J_K$ and $J_\perp$ represent the antiferromagnetic couplings for $f_A$-$\psi$ and $\psi$-$\psi^\prime$, respectively. 

In a fully self consistent calculation, the dynamics of $\psi^\prime$ would be explicitly solved. Here, we assume there is a ferromagnetic order at low temperature, then define the effective Zeeman field as $B=J_K\braket{S^z_{i;\psi}}$ and $B^\prime=-(J_K\braket{S^z_{i;A}}+J_\perp\braket{S^z_{i;\psi^\prime}})$. Since $J_K, J_\perp > 0$, $B$ and $B^\prime$ share the same sign. For the results presented in the main text, we adopt the phenomenological values $B=0.1$ and $B^\prime=0.2$.

We note that it is also possible that $\psi^\prime$ forms as a spin liquid state. In this case, the ancilla wavefunction takes the form as in Eq.~\ref{eq:ancilla_wavefunction}, written as:
\begin{equation}
    \ket{\Psi_\mathrm{ancilla}}=P_S\left(\ket{\mathrm{Slater}[c_B,f_A,\psi]}\otimes\ket{\Psi_{\psi^\prime}}\right),
\end{equation}
where $\ket{\Psi_{\psi^\prime}}$ is the spin liquid wavefunction of $\psi^\prime$.

\section{Exciton parameter in real space}

\begin{figure}[t]
    \centering
    \includegraphics[width=0.95\linewidth]{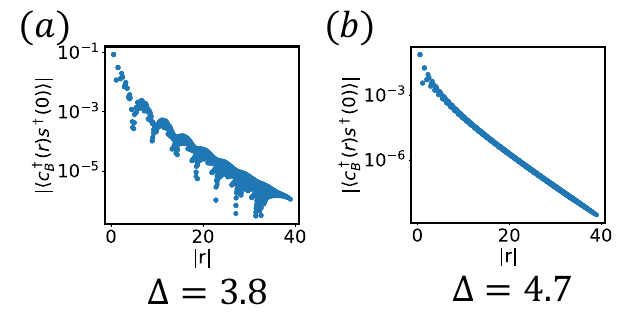}
    \caption{Real space exciton correlator $\langle c^\dagger_{B;\alpha;l}(r)s_{\bar{\alpha};\bar{l}}^\dagger(0)\rangle$ for $\Delta=3.8$ and $4.7$, showing oscillatory exponential decay in the BCS regime (a) and purely exponential decay in the BEC regime (b). }
    \label{fig:exciton_real}
\end{figure}

We provide the real space exciton order parameter $\langle c^\dagger_{B;\alpha;l}(r)s^\dagger_{\bar{\alpha};\bar{l}}(0)\rangle$ shown in Fig.~\ref{fig:exciton_real}(a)(b). For $\Delta=3.8$, the system is in the BCS regime, $\langle c^\dagger_{B;\alpha;l}(r)s_{\bar{\alpha};\bar{l}}^\dagger(0)\rangle$ shows an oscillatory exponential decay, which is a characteristic of a Fermi-surface-based pairing mechanism. In contrast, for $\Delta=4.7$, the system is in the BEC regime, $\langle c^\dagger_{B;\alpha;l}(r)s_{\bar{\alpha};\bar{l}}^\dagger(0)\rangle$ decays purely exponentially.  

\section{Tuning $U$}
\begin{figure}[t]
    \centering
    \includegraphics[width=0.95\linewidth]{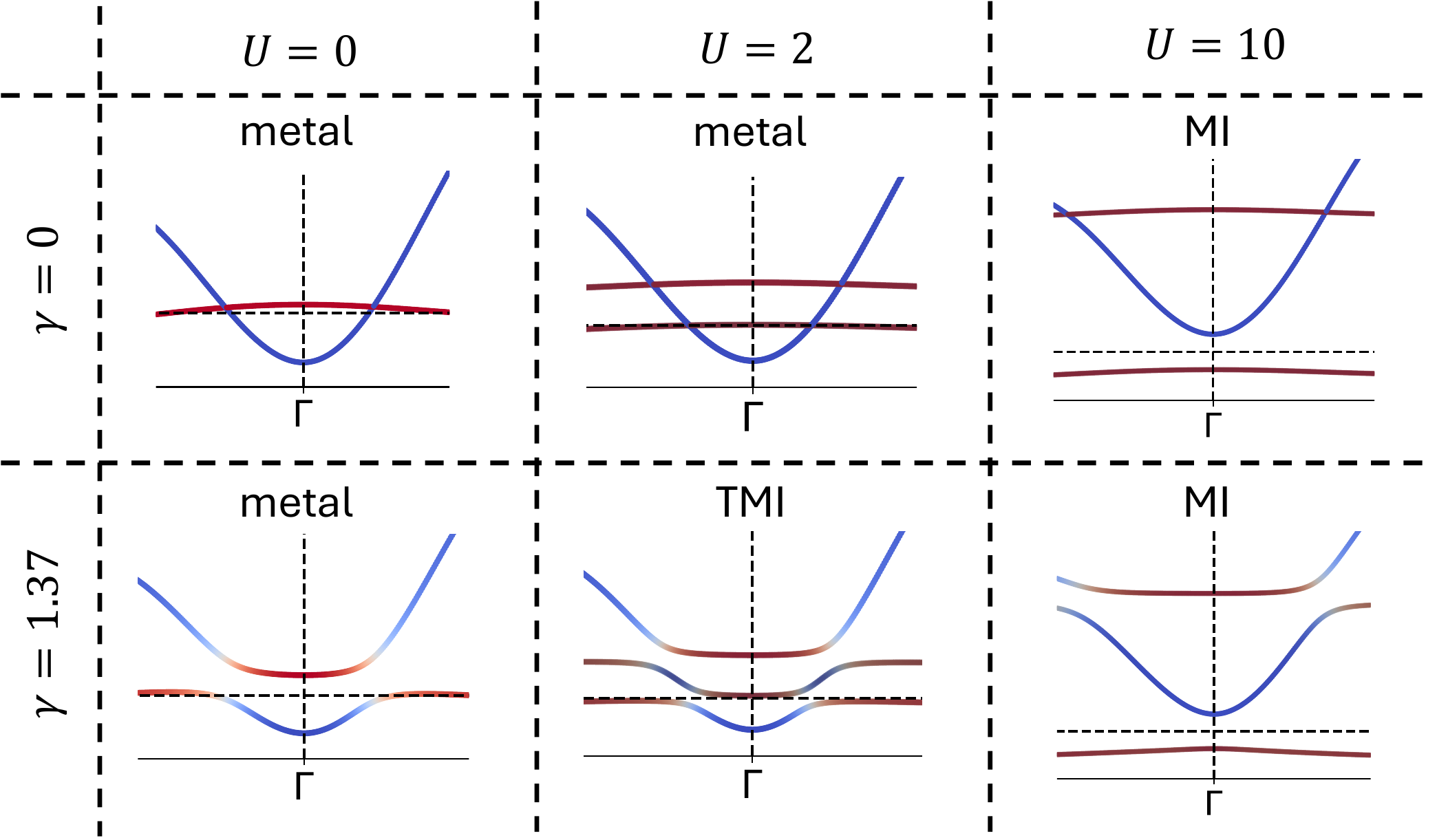}
    \caption{Parameters $\phi=0.78\pi,t_A=-0.1,t_B=2.5,\Delta=4.5$. Band structures per flavor of the effective Hubbard model for various interaction strengths $U$ and inter-sublattice hybridizations $\gamma$. Top row: $\gamma = 0$, bottom row: $\gamma = 1.37$. For small $U$, the system is metallic. At large $U$, a full Mott gap opens, leading to a trivial Mott insulator. For intermediate $U$ and finite $\gamma$, a symmetric topological Mott insulator emerges, characterized by band inversion near the $\Gamma$ point. This nontrivial topology originates from the interplay between inter-sublattice hybridization and on-site interaction, which reconstructs the Hubbard bands into a topologically nontrivial configuration. }
    \label{fig:tuneU}
\end{figure}
In this section, we provide the Hubbard bands per flavor by tuning interaction $U$ and inter-sublattice hybridization $\gamma$. The band structures per flavor are shown in Fig.~\ref{fig:tuneU}. For $\gamma = 0$ (top row), the system remains metallic for small $U$, and a Mott gap gradually opens only at large $U$, driving the system into a trivial Mott insulator. 

When a finite inter-sublattice hybridization $\gamma = 1.37$ is introduced (bottom row), the situation changes qualitatively: at intermediate interaction strength ($U = 2$), a topological Mott insulator emerges. In this phase, the band inversion of $A$ and $B$ sublattices gives rise to a non-zero Chern number for lower Hubbard bands. Away from the $\Gamma$ point, the dispersion resembles that of a trivial Mott insulator, dominated by the contribution from a single sublattice. Finally, for large $U$, the fluctuations on $B$ sublattice are suppressed. The system evolves into a trivial Mott insulator, characterized by a fully gapped, non-topological spectrum.
\end{document}